\documentclass[aps,prb,twocolumn,groupedaddress,showpacs]{revtex4}
\usepackage{graphicx}
\newcommand{\be}{\begin{equation}}
\newcommand{\ee}{\end{equation}}
\newcommand{\bea}{\begin{eqnarray}}
\newcommand{\eea}{\end{eqnarray}}

\begin{document}

\title{Numerical study of the frustrated ferromagnetic spin-$\frac{1}{2}$ chain  }

\author{S. Mahdavifar}

\address{Department of Physics, University of Guilan,
P.O.Box 41335-1914, Rasht, Iran}

\begin{abstract}

\leftskip 2cm \rightskip 2cm

The ground state phase diagram of the frustrated ferromagnetic spin-1/2
chain is investigated using the exact diagonalization technique.
It is shown that there is a jump in the spontaneous magnetization and the ground state of the system undergos to a phase transition from a ferromagnetic phase to a phase with dimer ordering between next-nearest-neighbor spins. Near the quantum transition point, the
critical behavior of the ground state energy is analyzed
numerically. Using a practical finite-size scaling approach, the
critical exponent of the ground state energy is computed. Our
numerical results are in good agreement with the results obtained
by other theoretical approaches.
\end{abstract}



\maketitle
\section{Introduction} \label{sec1}
  The physics of frustrated quantum spin systems have been
  attracted much interest from experimental and theoretical points
  of view. The spin-$\frac{1}{2}$ Hamiltonian of the frustrated
  model on a periodic chain of $N$-sites is
\begin{eqnarray}
H=\sum_{n=1}^{N}(J_{1}\overrightarrow{S}_{n}.\overrightarrow{S}_{n+1}+\
J_{2}\overrightarrow{S}_{n}.\overrightarrow{S}_{n+2}),\label{hamiltoni1}
\end{eqnarray}
where $\overrightarrow{S}_{n}$ represents the $S=\frac{1}{2}$
operator at the n-th site, and $J_{1}$, $J_{2}$ are the
nearest-neighbor (NN) and next-nearest-neighbor (NNN)
interactions. We introduce the parameter
$\alpha=\frac{J_{2}}{\mid J_{1}\mid }$ for convenience.

This model with NN and NNN antiferromagnetic interactions ($J_{1},
J_{2}>0$) is well studied\cite{haldane82, tonegawa87, nomura92,
bursill95, majumdar69, white96, japaridze98, jafari06}. This chain is well known to
display a quantum phase transition from a gapless, translationally
invariant state with algebraic spin correlations (the spin fluid
phase) to dimer gapful state at $\alpha_{c}\simeq
0.2411$\cite{nomura92}. At the Majumdar-Ghosh
point\cite{majumdar69}, i.e. at $\alpha=0.5$, the ground state is
exactly solvable. It is a doubly degenerate dimer product of
singlet pairs on neighboring sites. In general, the ground state
is doubly degenerate for $\alpha>\alpha_{c}$. For large $J_{2}$
($\alpha>0.5$) an incommensurate phase appears in the ground state
phase diagram\cite{bursill95, white96}. The behavior of frustrated chains in the presence of a uniform magnetic field was first studied by R. Chitra\cite{chitra97}. Recently the effect of a
uniform magnetic field on the $J_{1}-J_{2}$ model has been
discussed\cite{kolezuk05}. They have shown that a chiral phase
emerges in isotropic frustrated spin chains as well, if they are
subject to a strong external magnetic field. When $J_{1}>0$ and
$J_{2}<0$ (AF-F), the system is believed to be in a gapless
antiferromagnetic phase for any permissible values of $J_{1}$ and
$J_{2}$.

Relatively little attention has been paid to frustrated
ferromagnetic chains, i.e., $J_{1}<0$ and $J_{2}>0$. From
experimental point of view, the recent discovery of materials
are described by parameters with this combinations of signs. The
$Rb_{2}Cu_{2}Mo_{3}O_{12}$ is believed to be
described\cite{drechsler06, hase04} by $J_{1}\sim -3J_{2}$, and
$LiCuVo_{4}$ which lies in a different parameter regime with
$J_{1}\sim -0.3J_{2}$\cite{enderle05}. A recent
study\cite{heidrich06} of the thermodynamics of the model (1) was
motivated by the experimental results for
$Rb_{2}Cu_{2}Mo_{3}O_{12}$. From theoretical point of view, the
later model has been subject of many studies\cite{heidrich06,
tonegawa89, chubukov91, krivnov96, aligia01, lu06, jafari07}. The complete
picture of the phases of this model as a function of the
frustration parameter $\alpha$ is unclear up to now.

In the case of $J_{1}<0$ and $J_{2}>0$ (F-AF) with
$0\leq\alpha<\frac{1}{4}$, the ground state is fully ferromagnetic
and lies in the subspace $S_{tot}=N/2$ with the degeneracy $N+1$,
and becomes\cite{tonegawa89} an ($S=0$) incommensurate singlet state\cite{bader79, hamada89} 
for $\alpha>\frac{1}{4}$, also the lattice translational symmetry is
thought to be broken. It is suggested that in this incommensurate
singlet state, the gap is strongly suppresed\cite{itoi01}. At the critical
point $\alpha_{c}=\frac{1}{4}$, two distinct configurations with
the energy
\begin{eqnarray}
E_{g}=-\frac{3}{16}N \mid J_{1}\mid,\label{eg1}
\end{eqnarray}
are the ground states\cite{hamada88}. One is fully ferromagnetic
with $S_{tot}=N/2$, the other is a singlet state with $S_{tot}=0$.
The wave function of the singlet state at $\alpha_{c}=\frac{1}{4}$
is known exactly\cite{hamada88, dmitriev97}.

In the vicinity of the critical point $\alpha_{c}=\frac{1}{4}$, at
$0<\gamma\ll1$ ($\gamma=\alpha-\frac{1}{4}$) the singlet ground
state energy behaves as $E_{0}\sim \gamma^{\beta}$, where $\beta$
is a critical exponent. The classical approximation gives
$\beta=2$. The spin-wave theory as well as some other
approximations\cite{white96, krivnov96} do not change this
critical exponent. In Ref.[26], using variational approaches,
it has been shown that the quantum fluctuations definitely change the
classical critical exponent. They conjectured that strong
quantum fluctuations change the critical exponent and
$\beta=\frac{5}{3}$. In a recent work, Dmitriev
et.al.\cite{dmitriev07} have studied the properties of this model
using the perturbation theory (PT) in the small parameter
characterizing the deviation from the transition point. They
considered the Hamiltonian (1) as
\begin{eqnarray}
H&=&H_{0}+V_{\gamma}\nonumber \\
H_{0}&=&-\sum_{n}\overrightarrow{S}_{n}.\overrightarrow{S}_{n+1}+\frac{1}{4}\sum_{n}\overrightarrow{S}_{n}.\overrightarrow{S}_{n+2}\nonumber \\
V_{\gamma}&=&\gamma \sum_{n}\overrightarrow{S}_{n}.\overrightarrow{S}_{n+2},
\end{eqnarray}
with a small parameter $0<\gamma\ll 1$. Since the perturbation
$V_{\gamma}$ conserves the total spin $S^{2}$, the PT to the
lowest singlet state $\mid \psi\rangle$ of the Hamiltonian $H_{0}$
involves only singlet excited states. They showed that the PT
allow them to estimate the critical exponent of the ground state
energy as
\begin{eqnarray}
E_{0}(\gamma)\sim-N \gamma^{\beta}~~~~~\beta=5/3,\label{eg0}
\end{eqnarray}
which is in good agreement with their previous
result\cite{dmitriev06}. On the other hand, they have claimed that the exact diagonalization of finite chains shows a complicated iregular size dependence of the ground state energy, which makes the numerical estimation of the critical exponent $\beta$ impossible\cite{dmitriev07}. In a very recent work, the ground state
phase diagram of the spin-1/2 zigzag chain with weakly anisotropic
ferromagnetic NN and antiferromagnetic NNN interactions is
studied\cite{dmitrievarx}. It is shown that the ground state phase
diagram consists of the fully polarized ferromagnetic, the
commensurate spin-liquid and the incommensurate phases.

In this paper, we present our numerical results on the ground
state phase diagram of the 1D frustrated ferromagnetic spin-1/2 model.
Our results are obtained using the exact diagonalization
technique. In section 2, we present the results of exact diagonalization calculations on the ground state phase diagram of the model. In section 3, we discuss a practical finite-size scaling
approach and find the critical exponent of the ground state energy
in the vicinity of the critical point $\alpha_{c}=\frac{1}{4}$.
Finally, the summary and conclusions are presented in section 4.


\section{The ground state phase diagram}     \label{sec2}

An inportant goal in the study of quantum spin systems is the search for novel states emerging from competing interactions in the ground state phase diagram. In particular, the study of continuous phase transitions, has been one of the most
fertile branches of theoretical physics in the last decades. Each
phases can usually be characterized by an order parameter. Often,
the choice of an order parameter is obvious, however in some cases
finding an appropriate order parameter is complicated. As we mensioned, the complete
picture of the phases of this model as a function of the
frustration parameter $\alpha$ is not completely clear. It is known that the ground state is ferromagnetic at $0<\alpha<\alpha_{c}$ and a second order phase transition happens to the incommensurate singlet phase.

\begin{figure}[tbp]
\centerline{\includegraphics[width=8cm,angle=0]{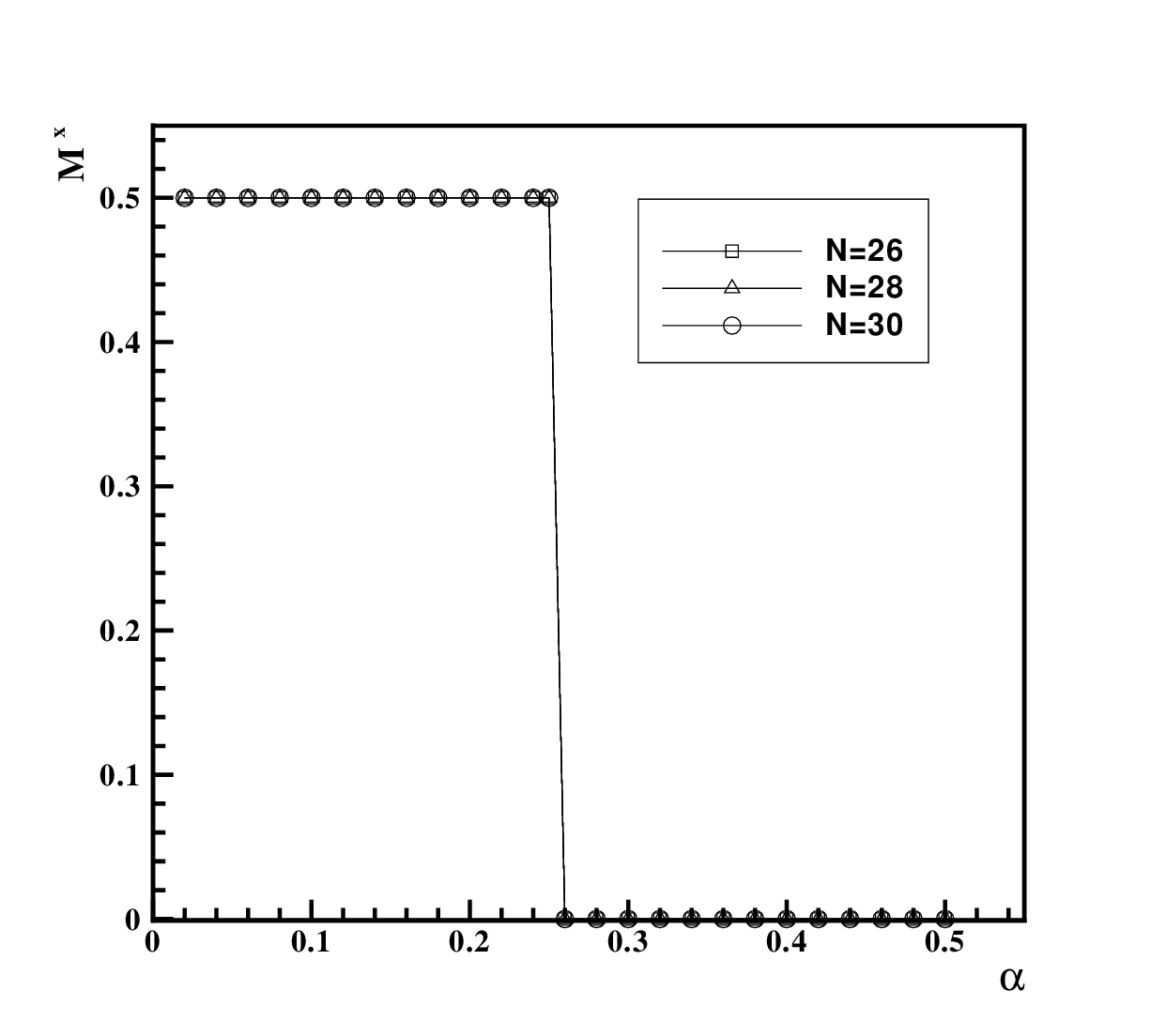}} \caption{
The spontaneous magnetization $M^x$ as a function of the parameter $\alpha$ for different chain lengths $N=26, 28, 30$.  } \label{fig1}
\end{figure}

In order to explore the nature of the spectrum and the phase transition, we used the Lanczos method to diagonalize numerically finite (up to $N=30$ sites) chain systems. The energies of the few lowest eigenstates were obtained for chains with periodic boundary conditions. The Lanczos method
and the related recursion methods,\cite{lanczos, haydock, grosso2,
lin1} possibly with appropriate implementations, have emerged as
one of the most important computational procedures, mainly when a
few extreme eigenvalues are desired. 

To recognize the different phases induced by the NNN exchange interaction in the ground state phase diagram, we have
implemented the Lanczos algorithm of finite size chains
to calculate the order parameters and the various
spin correlation functions.
The first insight into the nature of the different phases can be
obtained by studying the uniform magnetization 
\begin{eqnarray}
M^{x, y, z}=\frac{1}{N}\sum_{j}\left\langle  S_{j}^{x, y, z}\right\rangle,
\end{eqnarray}
where the notation $\langle...\rangle$ represent the expectation
value at the lowest energy state.

In Fig.~\ref{fig1} we have plotted the spontaneous magnetization, $M^{x}$ vs $\alpha$ for the chain of different lengths $N=26, 28, 30$. For arriving at this plot we considered
$\mid J_{1}\mid=1$ and different values of the parameter $0<\alpha<0.5$. One of the most interesting known properties of this model is that the magnetization as a function of applied magnetic field displays a jump for certain parameters\cite{gerhardt98, hirata00, aligia01}. It can be seen that the spontaneous magnetization $M^{x}$, remains close to the saturation value for $0<\alpha<\alpha_{c}$. This behavior is in agreement with expectations based on the general statement that for values of the parameter $0<\alpha<\alpha_{c}$, the ground state is in the gapped ferromagnetic phase. At a critical value $\alpha=\alpha_{c}$, the spontaneous magnetization jumps to zero. However, we observe that the metamagnetic phase transition occurs also in the absence of the external uniform magnetic field. The zero value of the spontaneous magnetization in the region $\alpha>\alpha_{c}$, shows that the ground state of the model is not magnetic.

\begin{figure}
\centerline{\includegraphics[width=8cm,angle=0]{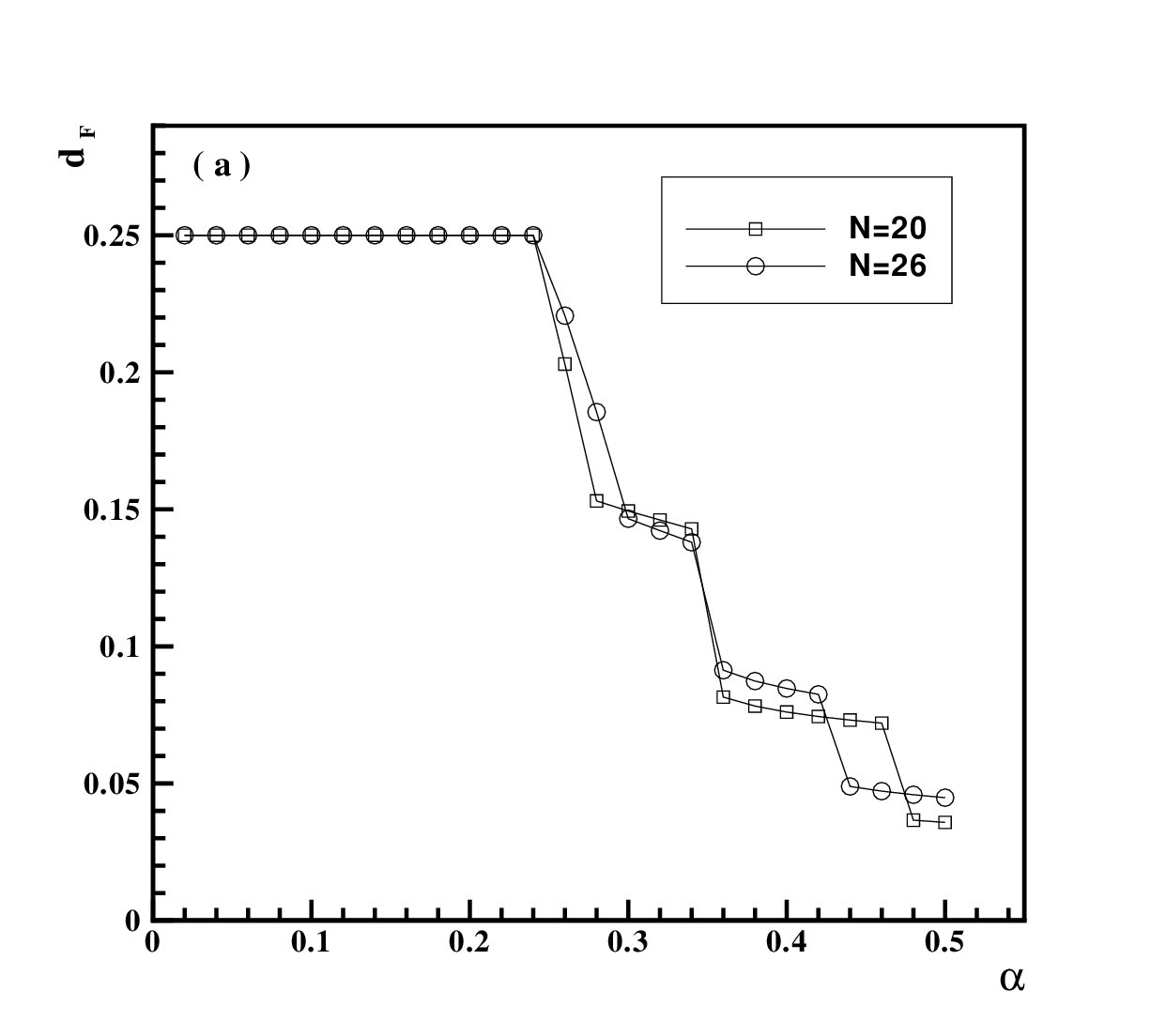}}
\centerline{\includegraphics[width=8cm,angle=0]{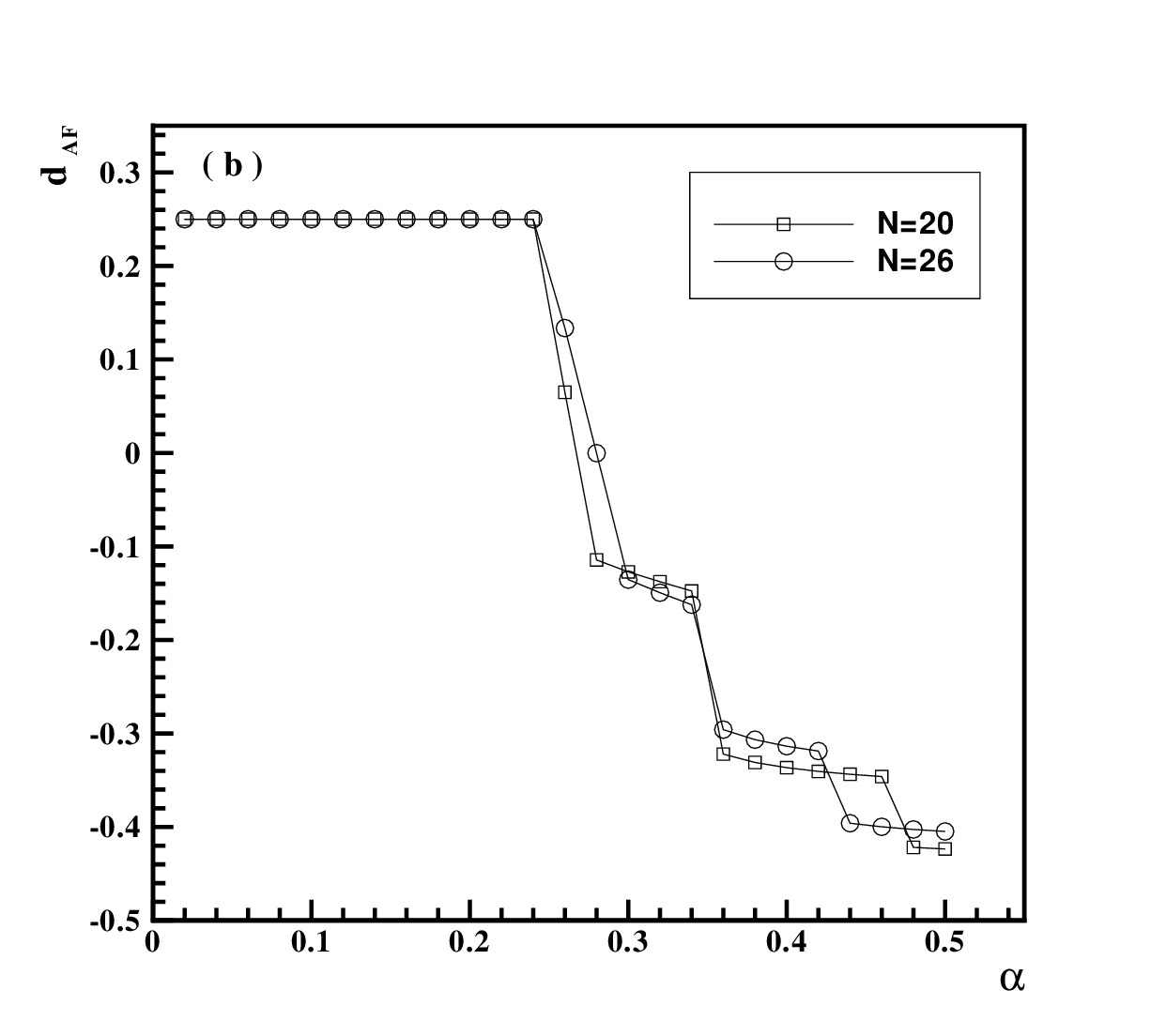}}
\caption{ (a) The F-dimer order parameter $d_F$ as a function of parameter $\alpha$ for different chain lenghts $N=20, 26$. (b) The AF-dimer order parameter $d_{AF}$ as a function of parameter $\alpha$ for different chain lenghts $N=20, 26$.} \label{fig2}
\end{figure}

To disply the ground state magnetic phase dagram of the model we have calculated the dimer order parameters. Because of two type of coupling constants, we introduce two kind of dimerization as
\begin{eqnarray}
d_F=\frac{1}{N}\sum_{n}\left\langle \overrightarrow{S}_{n}.\overrightarrow{S}_{n+1} \right\rangle,
\end{eqnarray}
\begin{eqnarray}
d_{AF}=\frac{1}{N}\sum_{n}\left\langle \overrightarrow{S}_{n}.\overrightarrow{S}_{n+2} \right\rangle.
\end{eqnarray}
It is clear that the parameter $d_{F}(d_{AF})$  is the F(AF)-dimer order parameter. In Fig.~\ref{fig2}a(b) we have plotted F(AF)-dimer order parameter as a function of the parameter $\alpha$ for the chain with $\mid J_{1}\mid=1.0$ and different values of the chain lengths $N=20, 26$. As is clearly seen from this figure, for $\alpha<\alpha_{c}$, $d_{F}$ and $d_{AF}$ are very close to $0.25$, which confirm that the ground state of the system is in the fully polarized ferromagnetic phase. For $\alpha>\alpha_{c}$ and enough large values of parameter $\alpha$, F-dimer order parameter is slightly more than zero $d_{F}\sim 0.05$ but the AF-dimer order parameter is less than saturation value ($-0.75$) $d_{AF}\sim -0.45$. Thus, increasing the antiferromagnetic exchang $J_2$ from critical value $\alpha_c$, quantum fluctuations suppress the ferromagnetic ordering and the system smoothly undergoes a transition from a ferromagnetic phase into a phase with the dimer ordering between the NNN spins. In this case of finite systems and with chosen values of the exchanges, due to the quantum fluctuations the values of the order parameters $d_F$ and $d_{AF}$ deviate from the classical values $0(-0.75)$ in the region $\alpha>\alpha_{c}$. The oscilations of $d_F$ and $d_{AF}$ (quasi-plateau's) at finite $N$ in the region $\alpha>\alpha_{c}$, are the result of level crossing between the ground state and excited states of the model. 

To obtain additional insight into the nature of different phases, we have also calculated the $x, y,$ and $z$ components of the dimer-order parameters. We have found that $d_{F}^{x}(d_{AF}^{x})$ is very close to the saturation value $0.25$ in the region $\alpha<\alpha_c$. The dimerization perpendicular to the $x$ axis remains small and close to zero in complete agreement with the magnetization results. As soon as the antiferromagnetic exchange $J_2$ increases from the critical value $\alpha_c$, $d_{F}^{x}(d_{AF}^{x})$ jumps to zero. However, all components of the F-dimer order parameter ($d_{F}^{x, y, z}$) remain close to zero in the region $\alpha>\alpha_c$. Which shows that there is no long-range ferromagnetic order in the region $\alpha>\alpha_c$. In contrast, components of the AF-dimer order parameter ($d_{AF}^{x, y, z}$) smothly change from almost zero to the value $-0.15$. Due to the quantum fluctuations induced by the ferromagnetic exchange $J_1$, the value of these components deviate from the saturation value $-0.25$.

Thus, our numerical results show that the ground state phase
diagram of the frustrated ferromagnetic spin-$\frac{1}{2}$ chain for small values of the antiferromagnetic exchange ($\alpha<0.5$) contains, besides the gapped ferromagnetic phase, the AF-dimer phase. Each phase is characterized by
its own type of long-range order: the ferromagnetic order along $x$ axis in the ferromagnetic phase; the AF-dimer order between NNN spins in the AF-dimer phase.


\section{the scaling behavior of the ground state energy \label{sec3}}

The finite size scaling method is a way of extracting values for
critical exponents by observing how measured quantities vary as
the size $L=N a$ (a is the lattice spacing and we will consider to be 
one) of the system studied changes. In fact, this method consists
of comparing a sequence of finite lattices. The finite lattice
systems are solved exactly, and various quantities can be
calculated as a function of the lattice size $L$, for small $L$.
Finally, these functions are scaled up to $L\longrightarrow
\infty$ \cite{okamoto86}. Two steps are needed before these ideas can be realized.
First, one needs a procedure for solving the finite lattice systems
exactly. Second, one needs a procedure for extrapolating from
finite to infinite $L$. In the step one, we have used the Lanczos method to obtain the ground  state energy. We also checked our numerical results by the modified 
Lanczos method\cite{grosso1}. Using the modified Lanczos method one can get the excited state energies at the same accuracy as the ground state one. We did not find any irregular size dependence of the ground state energy in our numerical results.     
In the following, we present
our finite-size scaling approach for the ground state energy.

Using Lanczos method, we can compute the ground state energy as a
function of the chain length $N$ and the parameter $\gamma$ as
$E_{0}(N,\gamma)$. We have implemented the modified Lanczos algorithm on finite size
chains ($N=10, 12, 14, ...,28$) by using periodic boundary conditions
to calculate the ground state energy as a function of the parameter $\gamma$.

In the case of $\gamma=0$, the spectrum of the 1D F-AF
$J_{1}-J_{2}$ model is gapless. The ground state energy in the
thermodynamic limit behaves as Eq.($\ref{eg1}$). By checking
the behavior of the function $E_{0}(N, \gamma=0)$ as a function of
$N^{\alpha}$ ($E_{0}=A N^{\alpha}$), found the best fit to
our data yielded $A=-0.1875$ and $\alpha=1.0$, which shows very
good agreement with the analytical result Eq.($\ref{eg1}$).

Now let us introduce our finite size scaling procedure to find the
correct critical exponent of the ground state energy in the
vicinity of the critical point $\alpha_{c}=\frac{1}{4}$. First, we
write the scaling function $f(x)$ as the following expression,
\begin{eqnarray}
N(E_{0}(N, \gamma)-E_{0}(N, \gamma=0))=f(x),\label{eg2}
\end{eqnarray}
where $x=N\gamma^{\beta}$ is a scaling parameter. As expected, the
behavior of this equation in the combined limit
\begin{equation}
N\longrightarrow\infty,~~~~~\gamma\longrightarrow0~~~~~(x\gg1)
\end{equation}
is consistent with Eq.($\ref{eg0}$). Thus it can be assumed that
the asymptotic form of the scaling function is
\begin{eqnarray}
f(x)\sim x^{\phi},
\end{eqnarray}
and the $\phi$-exponent in the large-$x$ regime ($x\gg 1$) must be
equal to one ($\phi=1$). Then we get in the large-$x$ regime
\begin{equation}
\lim_{N \to \infty (x \gg 1)} f(x)=N(E_{0}(N, \gamma)-E_{0}(N, 0)) \sim x. \label{ngap}
\end{equation}
This equation shows that the large-$x$ behavior of the scaling
function $f(x)$ is linear in $x=N\gamma^{\beta}$ where the scaling
exponent of the ground state energy is $\beta$. We should note
that in using the Lanczos method we are limited to consider the
maximum value of $N=30$\cite{saeed06}. Moreover, since the scaling behavior is restricted to the limit $\gamma\longrightarrow 0$, we should consider as soos as possible very small values of $\gamma<0.002$. Therefore, the value of $x$ cannot be increased in
this method. However, we are not allowed to read the scaling
exponent of the ground state energy which exists in the
thermodynamic limit ($N\longrightarrow\infty$ or $x\gg1$). Thus,
we have to find the scaling behavior from the small-$x$ regime.
According to our numerical computations where $N\leq30$, the small-$x$
regime is equivalent to very small values of the parameter
$\gamma$. In this case the ground state energy of the finite size
system basically represents the perturbative
behavior\cite{saeed06}
\begin{eqnarray}
E_{0}(N, \gamma)&=&B^{(0)}(N)+B^{(1)}(N)\gamma \nonumber \\
&+&B^{(2)}(N)\gamma^{2}+...\label{bast}
\end{eqnarray}
The effect of higher-order terms can be neglected for $\gamma\leq
0.002$ to a very good approximation. The first coefficient in the
perturbation expansion $B^{(0)}(N)$ is the same as $E_{0}(N,
\gamma=0)$.
\begin{figure}[tbp]
\centerline{\includegraphics[width=9cm,angle=0]{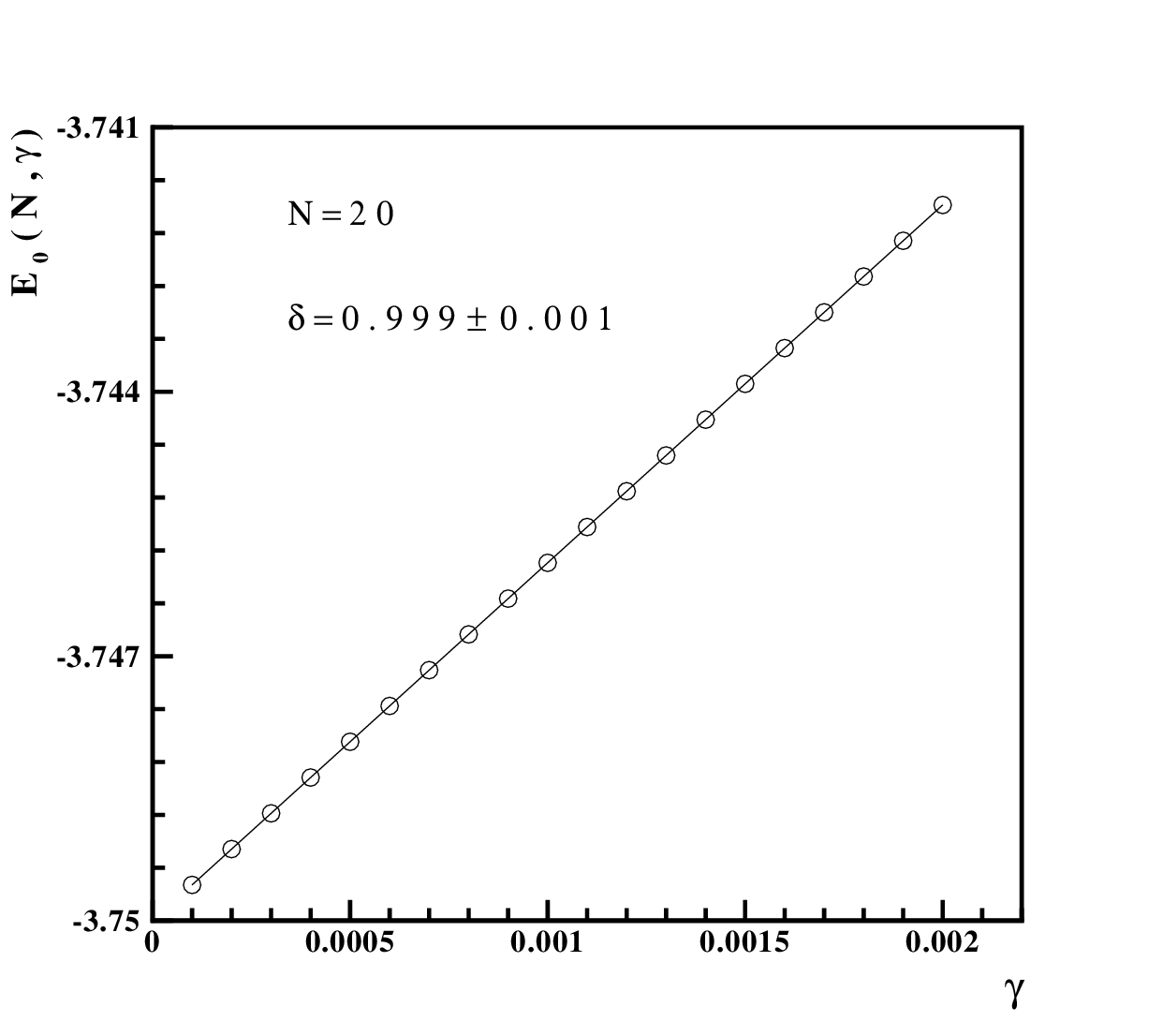}}
\caption{ The value of the ground state energy $E_{0}(N, \gamma)$
versus the parameter $\gamma$ close to critical point
$\gamma_{c}=0$. The results reported are for chain lengths $N=20$
and best fit is obtained by using equation ($E_{0}(N,
\gamma)\propto \gamma^{\delta}$) with $\delta=0.999\pm 0.001$. }
\label{fig3}
\end{figure}
\begin{figure}[tbp]
\centerline{\includegraphics[width=9cm,angle=0]{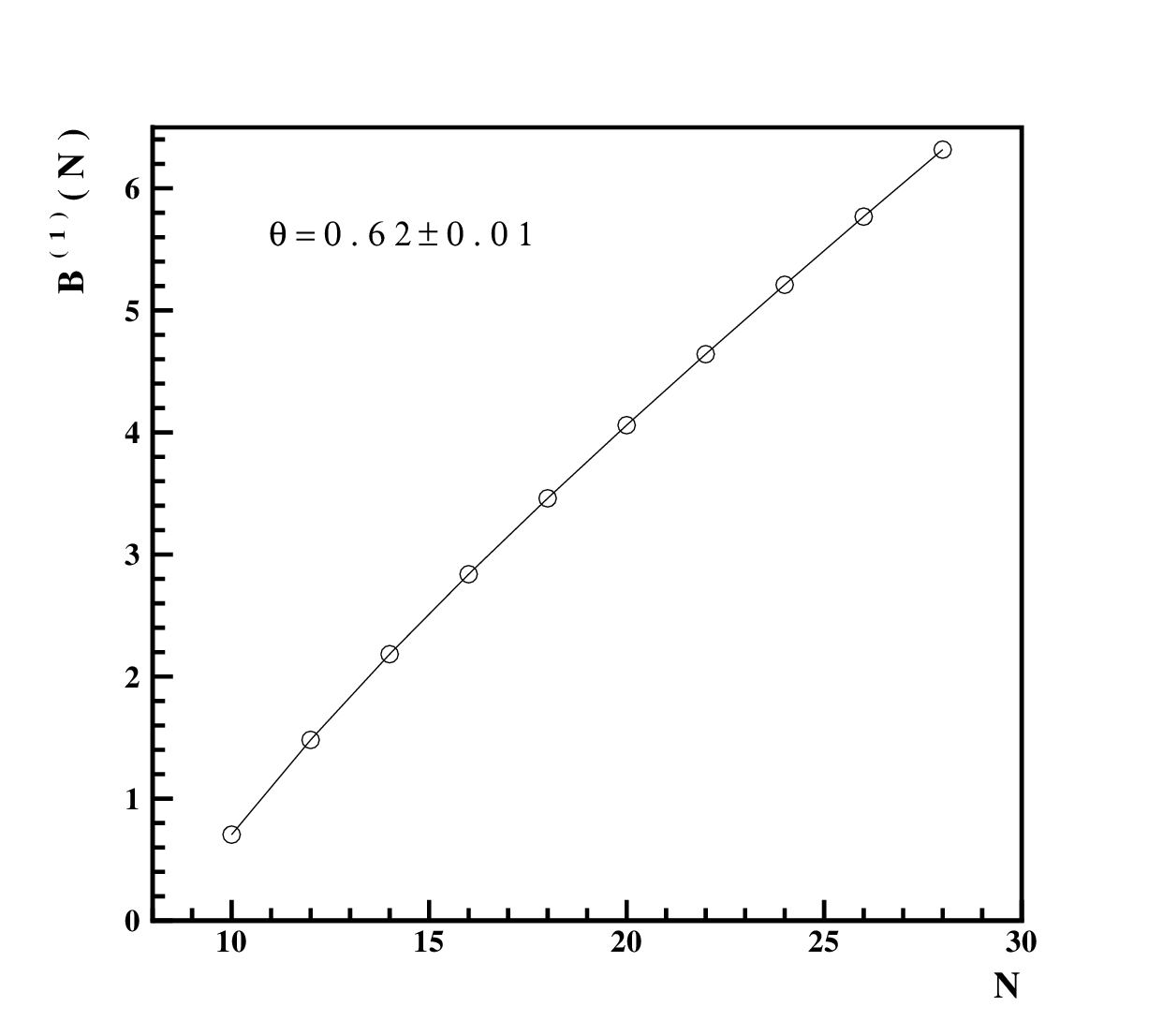}} \caption{
The value of the scaling function $B^{(1)}(N)$ versus the chain
length $N=10, 12, 14, ..., 28$. The best fit is obtained by using Eq.($\ref{am1}$)
with $\theta=0.62\pm 0.01$. } \label{fig4}
\end{figure}
To find a relation between other coefficients and correct critical
exponent of the ground state energy, it is more convenient rewrite
Eq.($\ref{eg2}$) as\cite{roomany80}
\begin{eqnarray}
E_{0}(N, \gamma)-E_{0}(N, 0) \sim g((N^{1/\beta})\gamma),
\end{eqnarray}
where $f(x)=Ng(x)$. This implies
\begin{eqnarray}
\frac{\partial^{m} E_{0}}{\partial \gamma^{m}}\mid_{\gamma_{c}}=N^{\frac{m}{\beta}} \times constant,
\end{eqnarray}
where $m$ is the order of the leading term in the perturbation
expansion. Using Eq.($\ref{bast}$) we obtain
\begin{eqnarray}
B^{(m)}(N)\propto N^{\frac{m}{\beta}}.
\end{eqnarray}
Now, if we consider the large-$N$ behavior of $B^{(m)}(N)$ as
\begin{equation}
\lim_{N\rightarrow\infty}B^{(m)}(N)\simeq a_{1} N^{\theta},
\label{am1}
\end{equation}
we find that the critical exponent of the ground state energy  is related to the
$\theta$-exponent as,
\begin{equation}
\beta=\frac{m}{\theta}. \label{am2}
\end{equation}

The above arguments suggest that we should look for the large-$N$ behavior of the coefficient
$B^{(m)}(N)$. To do this, in the first step we plotted
in Fig.~\ref{fig3} the ground state energy $E_{0}(N, \gamma)$ versus $\gamma$
$[0.0001\leq \gamma\leq 0.0002]$ for a fixed size $N=20$. The best fit to our data is
obtained with $\gamma=0.999\pm0.001$ ($E_{0}(N, \gamma)\propto
\gamma^{\delta}$), which shows that the first nonzero
correction in the perturbation expansion is the first-order ($m=1$).
We have also implemented our procedure for different values of the
sizes $N=10, 12, 14, ..., 28$ and found the same results for $m$ as we
expected.

In the second step, we fitted the results of the ground state
energy $E_{0}(N, \gamma)$ to the polynomials for $\gamma$ close to
$\gamma=0$ as Eq.($\ref{bast}$) up to $m=1$. Using this procedure
we found the coefficient of the first-order correction
perturbation, $B^{(1)}(N)$, as a function of $N$. Then we plotted
in Fig.~\ref{fig4} the function $B^{(1)}(N)$ versus $N$. The results have
been plotted for different sizes $N=10, 12, 14, ..., 28$ to derive
the $\theta$-exponent defined in Eq.($\ref{am1}$). We found the
best fit data for $\theta=0.62\pm 0.01$. Therefore, using
Eq.($\ref{am2}$) we have computed the ground state energy exponent
$\beta=1.61\pm0.01$. Our numerical results show very good
agreement with the exponent derived in the theoretical point of
view, Eq.($\ref{eg0}$).


\section{summary \label{sec5}}

To summarize, we have studied the ground state phase diagram of the frustrated ferromagnetic spin-1/2 chain for small values of parameter $\alpha<0.5$. We have implemented the Lanczos method to obtain the ground
state energy in small chains. The modified Lanczos method\cite{grosso2} is used also for checking the numerical results. Using the exact diagonalization results, we have calculated the various order parameters and spin structure factors as a function of the parameter $\alpha$. It is found that the spontaneous magnetization jumps to zero at the critical value $\alpha=\alpha_c=1/4$. Increasing the antiferromagnetic exchange $J_2$ from critical value $\alpha_c$, the system smoothly undergoes a transition from a ferromagnetic phase into a phase with the dimer ordering between the NNN spins.  
 
On the other hand, it is
believed that the ground state energy behaves as $E_{0} \sim
\gamma^{\beta}$, where $\beta$ is a critical exponent. From the
classical approximations and the spin-wave theory, it had been
obtained $\beta=2$. Using the variational approaches and
perturbation theory, it had been shown that the quantum
fluctuations definitely change the critical exponent and
$\beta=5/3$. On the other hand, it had been believed that the exact diagonalization of finite chains shows a complicated iregular size dependence of the ground state energy, which makes the numerical estimation of the critical exponent $\beta$ impossible\cite{dmitriev07}. 

In this paper, we have used the finite-size scaling approach to
investigate the critical exponent of the ground state energy.  To estimate the critical exponent of the ground state energy, we have introduced a proper scaling function $f(x)$ as Eq.($\ref{eg2}$). The scaling variable is defined as $x=N \gamma^{\beta}$. According to our approach the right scaling exponent of the ground state energy gives a linear behavior of the scaling function $f(x)$ versus $x$ for large $x$. But, the Lanczos numerical results are not able to get the large-$x$ behavior. 

 To find the correct critical
exponent of the ground state energy in the small-$x$ regime ($x\ll
1$), we have plotted the best fit to the data of the scaling
function $B^{(m)}(N)$, which is the coefficient of the $m$-order
perturbation expansion of the ground state energy. The critical
exponent of the ground state energy is computed with the relation
between the divergence of the leading term ($B^{(m)}(N)$) in the
perturbation expansion and the scaling behavior of the ground
state energy (Eq.($\ref{am2}$)). Our numerical results, confirm
that the quantum fluctuations are very important and change the critical exponent from
the classical value and $\beta=1.61\pm0.02$, in good agreement
with the analytical results(Eq.($\ref{eg0}$)).
\section{Acknowledgments}

I would like to thank G. I. Japaridze and T. Vekua for
insightful comments and fruitful discussions that led to an
improvement of this work.


\end{document}